\documentstyle[aps,preprint]{revtex}

\begin{document}
\begin{titlepage}
\begin{flushright}
{\tt hep-th/9907131}\\
{\tt  IMSc 99/07/25}~~ \\
Jan 2000 (ver 3)
\end{flushright}
\vfill
\begin{center}
{\Large \bf Worldsheet approaches to D-branes on
supersymmetric cycles}\\[1cm]
Suresh Govindarajan\footnote{Email: suresh@chaos.iitm.ernet.in}\\
{\em Department of Physics,\\ Indian Institute of Technology, Madras,\\
Chennai 600 036, India\\[10pt]}
T. Jayaraman and Tapobrata
Sarkar\footnote{Email: jayaram,sarkar@imsc.ernet.in}\\
{\em The Institute of Mathematical Sciences, \\ Chennai 600 113, India}
\end{center}
\vspace{2cm}
\begin{abstract}
We consider D-branes wrapped around supersymmetric cycles of Calabi-Yau
manifolds from the viewpoint of $N=2$ Landau-Ginzburg models with boundary
as well as by consideration of boundary states in the corresponding
Gepner models. The Landau-Ginzburg approach enables us to provide a target
space interpretation for the boundary states. The boundary states are
obtained by applying Cardy's procedure to combinations of characters 
in the Gepner models which are invariant under spectral 
flow. We are able to relate the two descriptions using the common 
discrete symmetries of the two descriptions. We thus provide 
an extension to the boundary, the bulk correspondence between
Landau-Ginzburg orbifolds and the corresponding Gepner models.
\end{abstract}
\vfill
\end{titlepage}

\section{Introduction}

  Dirichlet branes (D-branes), which are a simple realisation of
Ramond-Ramond charged solitonic objects in superstring theory, have played
an important role in our understanding of non-perturbative aspects of
string theory\cite{polchinski}. Among other things, they has played an
important role in the understanding of string duality and its relationship
with M-theory, analysis of stringy black holes, etc. D-branes as
formulated in string theory admit a world-sheet conformal field theory
(CFT) description in contrast to the description of these objects as
solitonic p-branes encountered as solutions of the low-energy effective
actions of string theory. 

  The world-volume theories of D-branes have provided us with interesting
examples of supersymmetric gauge theories. In its simplest form, the
world-volume spectrum of a flat D-brane is obtained by the dimensional
reduction of ten-dimensional super Yang-Mills theory. One may also
consider situations where some or all of the spatial directions of the
D-brane are wrapped around some cycle in a curved manifold that forms part
of ten-dimensional space-time. In order to preserve some supersymmetry it
has to be ensured that the cycle is actually a supersymmetric
cycle\cite{bbs,bsva,bsvb}.  In general, the bulk of the effort thus far
has been in understanding the cases where the curved manifold is obtained
by a reasonably simple modification of a flat manifold including that of
tori, and orbifolds of both tori and flat space. 

  However some progress has been made in providing a conformal field
theory description of D-branes wrapped around supersymmetric cycles in
Calabi-Yau spaces. The first important step was provided by the work of
Ooguri, Oz and Yin\cite{ooy}, who formulated the general boundary
conditions on the world-sheet $N=2$ super conformal field theory (SCFT)
that would be necessary to describe such cycles.  Subsequently using the
work of Cardy on boundary CFT\cite{cardy}, Recknagel and
Schomerus\cite{reck} described in some generality the boundary states in
the so-called Gepner models\cite{ictpgepner,gepner}, that would be
relevant to the description of both even and odd dimensional
supersymmetric cycles in the corresponding Calabi-Yau manifolds. Further
in refs. \cite{gutsat} some applications of this construction have been
pursued. A more general construction of boundary states relevant to
curved D-branes have also been pursued by Fuchs and Schweigert\cite{fs}.
In later work, Recknagel and Schomerus have also studied the role
of boundary operators in such constructions\cite{reckb}. Other approaches
have studied the case of D-branes in the context of group manifolds as
described by WZW models\cite{stanciua,stanciub,kato}.  While this paper
was in preparation, the important work of Brunner, Douglas, Lawrence and
R\"omelsberger\cite{quintic} appeared that studied in detail the structure
and several aspects of D-branes on the quintic, using both Gepner models
and other techniques. We consider the techniques of this paper to be
complementary to the ideas and results contained therein. 

  In this paper, we pursue two different worldsheet approaches to
understanding such D-branes wrapped on supersymmetric cycles in Calabi-Yau
manifolds. For simplicity we restrict ourselves to the case where all the
spatial coordinates are wrapped on the appropriate supersymmetric cycle
and hence from the viewpoint of the non-compact spacetime, we have a
zero-brane. From this point of view, the world-volume theory describes the
moduli of the corresponding D-brane wrapped on the cycle inside the
Calabi-Yau manifold\cite{bsva,bsvb}. 

 The two approaches that we use are the boundary $N=2$ supersymmetric
Landau-Ginzburg (LG) formulation and a boundary state construction in
terms of the Gepner model.  The Landau-Ginzburg formulation of strings on
Calabi-Yau manifolds has been very successful in understanding various
aspects of such closed string theories. We extend this by considering the
same LG models on worldsheets with boundary, in a manner that preserves a
$N=2$ worldsheet supersymmetry on the boundary. We find that these LG
models with boundary provide a natural description of D-branes wrapped on
both even and middle-dimensional supersymmetric cycles in the general
Calabi-Yau manifold. 

 In the second description, we use the Gepner model construction. However,
in contrast to other approaches mentioned earlier we consider linear
combinations of characters of the spacetime SCFT and the internal SCFT
that are invariant under spectral flow. With this approach we are able to
construct the cylinder partition functions in a manner that explicitly
demonstrates that some of spacetime supersymmetry is preserved and thus
leads to a vanishing partition function.  The associated boundary states
are constructed for these partition functions. This is illustrated by the
$1^3$ and $2^2$ Gepner models that describe a $T^2$ compactification. The
approach that we use is closely related to the techniques that have been
used by the Rome group to describe the construction of Type-I strings
using Gepner models\cite{romea,romeab,romeb,romec}. However we note that
their formulation has not kept track of the construction of D-brane states
and their properties even though they have investigated the question of
anomaly cancellation and tadpole cancellation in some
detail\cite{romea,romeab,romeb,romec}.  We are able to relate the boundary
state construction to the boundary condition LG description by making use
of a common discrete symmetry group occurring in both the Gepner model and
its corresponding LG orbifold.

	We would like to emphasise the point of view taken throughout
this paper. When we refer to D-branes, we mean the object which the
boundary CFT describes. In some cases, this coincides with the
supergravity description of D-branes. However, this is not expected to
be generically true especially in the case of compactifications on
Calabi-Yau three-folds where the ``large volume branes'' (these typically
admit supergravity descriptions) may exhibit rather different behaviour
in the ``small volume'' limit. According to the modified geometric hypothesis of
Douglas et al.\cite{quintic}, for the so called A-branes, i.e., D-branes which 
correspond
to A-type boundary conditions, the ``large volume'' results for central
charges and masses of A-branes do not get modified in the ``small
volume'' limit. This is not so for the case of B-branes, i.e., D-branes
which correspond to B-type boundary conditions. Based on this,
the geometric properties which we have extracted for the A-branes
in section 4, which
we construct using the boundary state formalism, is expected to hold.

The paper is organised as follows. Section 2 contains the necessary
background and the notation followed in the paper. In section 3, we
consider Landau-Ginzburg theories in the presence of boundaries which
preserve $N=2$ supersymmetry. We obtain general boundary conditions on the
LG fields when $N=2$ supersymmetry is preserved and relate these to
supersymmetric cycles.  In section 4, we study D-branes in Gepner models
making use of spectral flow invariant orbits. Cardy's prescription is used
after suitably resolving the S-matrix. The $1^3$ and $2^2$ models are
analysed in detail though the analysis is more general. We also attempt to
use the LG formulation in order to obtain a spacetime picture for the
boundary states. In section 5, we conclude with a discussion on open
issues and possible extensions of the work described in the paper. In an
appendix, we discuss the transformations of the boundary states in the
$2^2$ Gepner model under the action of the discrete symmetry group of the
model. 

\section{Background}
\subsection{ The $N=2$ Supersymmetry Algebra}

The generators of the $N=2$ super conformal algebra are the energy
momentum tensor, $T(z)$ its worldsheet superpartners, $G$  and
$\overline{G}$ of
conformal weight $3/2$, and a U(1) current, $J$, with conformal weight
$1$.  The algebra is given by the following relations that can be derived
from the operator product expansions of the generators\cite{warnerreview}. 
\begin{eqnarray}
\left[L_m, L_n\right]&=& \left(m-n\right)L_{m+n}~+~{c \over 12}m
\left(m^2-1\right)\delta_{m+n,0} \nonumber \\
\left[L_m, G_{n + a}\right] &=&\left( {m \over 2} - (n 
+ a)\right) G_{m+n +a} \nonumber \\
\left[L_m, \overline{G}_{n - a}\right] &=&\left( {m \over 2} - (n 
- a)\right) \overline{G}_{m+n - a} \nonumber \\
\left[J_m, J_n \right]&=& {c \over 3}m \delta_{m+n,0} \nonumber \\
\left[L_n, J_m \right]&=&-m J_{m+n} \nonumber \\
\left[J_n, G_{m + a}\right] &=&  G_{m+n + a}
\nonumber \\
\left[J_n, \overline{G}_{m - a}\right] &=& - \overline{G}_{m+n - a}
\nonumber \\
\{ G_{n+a}, \overline{G}_{m-a} \}&=&2L_{m+n}+\left(n-m+2a 
\right) J_{m+n}+{c \over 3} \left( (n+a)^2 - {1 \over 4} \right)
\delta_{m+n,0} 
\end{eqnarray}
The parameter $a\in[0,1)$. $a=0$ corresponds to the Ramond (R) algebra and
$a={1\over2}$ corresponds to the Neveu-Schwarz (NS) algebra. We shall
refer to states in representations of the Ramond algebra as Ramond states
and similarly, one obtains Neveu-Schwarz states. Primary states of the
$N=2$ algebra are labelled by their dimension $h$ and $U(1)$ charge $q$. 

A subset of the primary fields of the NS algebra are the chiral primary
fields.  which create states that are annihilated by the operator
$G_{-1/2}$, i.e,
\begin{equation}
G_{-1/2}| \phi \rangle = 0\quad,
\end{equation}
where $|\phi\rangle$ is the state created by a chiral primary field
$\phi$. The dimension and $U(1)$ charge of a chiral primary field satisfy
\begin{equation}
h_{\phi} = {q_{\phi} \over 2}\quad.
\end{equation}
Anti-chiral fields creates states annihilated by $\overline{G}_{-1/2}$ with
$h=-q/2$.

In a theory with $(2,2)$ worldsheet supersymmetry, i.e., theories with
$N=2$ supersymmetry in the holomorphic(left-moving)
 and anti-holomorphic(right-moving) sectors, one can construct four
combinations of the chiral and anti-chiral fields. These are $(c,c),
(a,a), (c,a),$ and $(a,c)$ states in the theory. 

An important aspect of the $N=2$ algebra is the existence of a spectral
flow isomorphism. One can show that the after the following redefinition: 
\begin{eqnarray}
L_n'~&=&~L_n~+~\eta J_n ~+~ {1 \over 6}\eta^{2}c \delta_{n,0} \nonumber 
\\
J_n'~&=&~J_n~+~{1 \over 3} \eta c \delta_{n,0} \nonumber \\
\left(G_r\right)'&=~&G_{r + \eta}\quad, \nonumber \\
\left(\overline{G}_r\right)'&=~&\overline{G}_{r - \eta}\quad,
\end{eqnarray}
the redefined operators also satisfy the $N=2$ algebra with a moding
shifted by the parameter $\eta$ ($a\rightarrow a+\eta$). This
correspondence can be carried over to the states in the representation of
the algebra. This is done by means of the spectral flow operator $U_\eta$
defined by
\begin{equation}
U_{\eta}~=~e^{i \sqrt{{c \over 3}} \eta \phi}\quad,
\end{equation}
where the $U(1)$ current is given by $J=i\sqrt{c/3}\ \partial_z \phi$. The
dimension and $U(1)$ charge of the new field obtained by spectral flow of
a primary field with weight $h$ and $U(1)$ charge $q$ is given by
\begin{eqnarray}
h_{\eta}&=&h  + q \eta + {\eta^2 \over 6}c \nonumber \\
q_\eta&=&q + \eta {c \over 3}
\end{eqnarray}
When $\eta={1\over2}$, the spectral flow operator interpolates between the
Neveu-Schwarz and the Ramond sectors. In the context of spacetime
supersymmetric string theory, this spectral flow relates spacetime bosons
to spacetime fermions.

For a given representation $p$ of the $N=2$ algebra, the character is
defined as
\begin{equation}
\chi_p\left(\tau, z, u \right)~=~e^{-2 i \pi u}\ {\rm Tr}\ [e^{2 i \pi z 
J_0}\ e^{2 i \pi \tau (L_0-{c \over 24})}]
\end{equation}
where the trace runs over the particular representation denoted by $p$ and
$u$ is an arbitrary phase. The explicit formulae for the characters of
certain models in terms of the Jacobi theta functions will be written down
later. Under spectral flow with parameter $\eta$, the character for the
$\eta$-shifted representation is given by
\begin{eqnarray}
\chi^{\eta}_p\left(\tau, z,0\right)&=&{\rm Tr}\ [e^{2 i \pi J_0'}\
e^{2 i \pi \tau (L_0' - {c \over 24})}] \nonumber \\
&=&\chi_p\left(\tau, z+\eta \tau, -{1 \over 6} \eta^2 \tau c 
- {1 \over 3} \eta z c \right) 
\label{twistedcharacter}
\end{eqnarray}

\subsection{Boundary states for N=2 SCFT }

A BPS state such as a D-brane wrapped on a supersymmetric cycle will
preserve half the spacetime supersymmetry. Using the correspondence
between spacetime supersymmetry and the existence of a global $N=2$
supersymmetry on the worldsheet, the presence of a BPS state will be
signalled by the boundary preserving a linear combination of the $(2,2)$
worldsheet supersymmetry. The analysis of Ooguri et al. shows that there
are two possible linear combinations\cite{ooy}\footnote{We denote
left-movers by the subscript $+$ and right-movers by the subscript $-$.}.\\
{\bf A-type boundary condition:}
\begin{equation}
J_+ = - J_-\quad,\quad G_+=\pm \overline{G}_-\quad,\quad e^{i\phi_+} =
e^{-i\phi_-}
\end{equation}
{\bf B-type boundary condition:}
\begin{equation}
J_+ = J_-\quad,\quad G_+ = \pm G_-\quad,\quad e^{i\phi_+} = (\pm)^d
e^{i\theta} e^{i\phi_-}\quad,
\end{equation}
where the $\phi_\pm$ are the scalars associated with the bosonisation of the
$U(1)$ current of the $N=2$ supersymmetry algebra in the left and
right-moving sectors. These boundary conditions are for the open string
channel. 

Boundary states which preserve a $N=2$ supersymmetry are expected to be
related to D-branes wrapping around supersymmetric cycles. The boundary
states satisfy the closed string equivalent of the above boundary
conditions.  In order to do this, we write the boundary conditions in the
closed string channel with the replacement $J_- \rightarrow -J_-$,
$G_- \rightarrow i G_-$ and
$\overline{G}_- \rightarrow i \overline{G}_-$ as compared to the open string
channel. The A-type boundary condition then reads,
\begin{equation}
\left(J_+ - J_-\right)|B\rangle=0\quad;\quad\left( G_+\pm
i\overline{G}_-
\right)|B\rangle = 0\quad,
\end{equation}
where $|B\rangle$ is a boundary state. The condition on the $U(1)$ current
picks out a selection rule for the fields of the theory that can
contribute to the boundary state, namely for the A-type boundary
condition, corresponding to D-branes wrapping around middle dimensional
cycles, we have $q_+ = q_-$ for the $U(1)$ charge. Thus, the $(c,c)$ and
$(a,a)$ states can contribute to the A-type boundary state while the
$(a,c)$ and $(c,a)$ states cannot. Similarly, for the B-type boundary
condition
\begin{equation}
\left(J_+ + J_-\right)|B\rangle = 0\quad;\quad\left( G_+\pm iG_-
\right)|B\rangle = 0
\end{equation}
implying that the $(c,a)$ and $(a,c)$ states
contribute to the boundary state.

Generalising a procedure due to Ishibashi, one can construct solutions of
the above conditions for all primary fields which satisfy the 
condition involving the two $U(1)$ charges in addition to the condition
on the conformal weights\cite{ishibashi}.  The explicit form 
of the Ishibashi state
associated with such a representation $a$ is given by
\begin{equation}
|a\rangle\rangle ~=~\sum_N |a;N\rangle \otimes U \overline{ |a;N\rangle}
\end{equation}
where $|a,N\rangle$ is an orthonormal basis for the representation $a$ and
$U$ is an anti-unitary matrix which preserves the highest weight state
$|a\rangle$. For A-type boundary conditions, one has to replace $U$ with
$U\Omega$ where $\Omega$ is the mirror automorphism of the $N=2$
algebra\cite{reck}. We shall label the Ishibashi states for the A-type and
B-type boundary conditions by $|a\rangle\rangle_A$ and
$|a\rangle\rangle_B$ respectively.

\subsection{Cardy's construction}

The set of Ishibashi states form a basis for the boundary states. Thus,
any boundary state $|\alpha\rangle$ is given by a linear combination of
the Ishibashi states
\begin{equation}
|\alpha\rangle = \sum_a\ {{\psi_\alpha}^a\over{({S_0}^a})^{1\over2}}\ |a\rangle\rangle\quad,
\end{equation}
where $S$ is the modular S-matrix and $0$ refers to the identity operator.
The ${\psi_\alpha}^a$ are not arbitrary but will have to satisfy a
consistency condition which we will now derive. The arguments are due to
Cardy\cite{cardy} but we will follow the discussion in ref. \cite{zuber}.
Consider a conformal field theory associated with a chiral algebra on a
cylinder with perimeter $T$ and length $L$ subject to boundary conditions
$\alpha$ and $\beta$. The partition function of the system can be
calculated in two ways: One can consider the result as coming from
periodic `time' $T$ evolution with the prescribed boundary conditions.
Topologically, this corresponds to an annulus. The annulus partition
function is given by
\begin{equation}
{\cal A}_{\alpha\beta} = \sum_i {n_{i\alpha}}^{\beta}\ \chi_i(q)\quad,
\label{annone}
\end{equation}
where ${n_{i\alpha}}^{\beta}$ denotes the number of times the irreducible
representation $i$ occurs in the spectrum of the Hamiltonian
$H_{\alpha\beta}$ (which generates the `time' evolution) and $q=e^{-\pi
T/L}$. Another way corresponds to treating the $L$ direction as time and
the partition function for time evolution from the boundary state
$|\alpha\rangle$ to the boundary state $|\beta\rangle$ is given by
\begin{equation}
{\cal C}_{\alpha\beta}=\sum_a {{{\psi_\alpha}^a({\psi_\beta}^a)^{\dagger}
\ \chi_a(\tilde{q})}\over{S_0}^a} \quad,
\label{cylone}
\end{equation}
where $\tilde{q}=e^{-4\pi L/T}$ and the sum is over Ishibashi states. On
equating eqn. (\ref{annone}) to the modular transformation $\tau
\rightarrow -1/\tau$ (with $\tau = i2L/T$) of eqn. (\ref{cylone}), one
obtains the following consistency condition: 
\begin{equation}
{n_{i\alpha}}^{\beta} =\sum_a {{S_i}^a \over {S_0}^b}
\ {\psi_\alpha}^a\ ({\psi_\beta}^a)^{\dagger}\quad.
\label{consist}
\end{equation}
In the above, note that the sum is over Ishibashi states while the index
$i$ is over characters of all irreducible representations of the chiral
algebra.  Note that these two are not necessarily the same except for
theories such as the one whose
toroidal partition function is given by the charge
conjugation modular invariant combination.

It can be shown\cite{zuber} that the matrices $n_i =
{(n_{i})_\alpha}^{\beta}$ form a representation to the fusion algebra
\begin{equation}
\sum_\beta {n_{i\alpha}}^{\beta} {n_{j\beta}}^{\gamma} =
\sum_k {N_{ij}}^k {n_{k\alpha}}^{\gamma}\quad,
\end{equation}
where ${N_{ij}}^k$ is the fusion matrix. In general, the boundary theory 
need not preserve all the symmetries in the bulk. More general situations
have been studied by Fuchs and Schweigert\cite{fstwo}. A simple example
which illustrates the general situation
is the three-state Potts model\cite{potts}.

Cardy has provided a solution to the consistency equation (\ref{consist})
for theories whose toroidal partition function is given by the
charge conjugation invariant.
He constructs boundary states (and hence boundary conditions)
corresponding to the representations $a$ which appear in the Ishibashi
states. Let us label the corresponding boundary states by
$|\tilde{a}\rangle$ given by
\begin{equation}
|\tilde{a}\rangle = \sum_b {{S_a}^b\over{({S_0}^b)^{1/2}}}\ |b\rangle\rangle
\quad,
\end{equation}
where the sum is over Ishibashi states. 
This solves eqn. (\ref{consist}) for 
\begin{equation}
{n_{i\tilde{a}}}^{\tilde{b}} = {N_{ia}}^b\quad,
\end{equation}
The consistency condition now turns into the Verlinde formula.

Complications can arise in attempting to apply Cardy's results directly.
One which we will encounter is that different representations may have the
same Virasoro character. This will show up as a multiplicity in the
appearance of the characters in the toroidal partition function. In
addition, the S-matrix will not have several of its usual properties such
as it being symmetric and so on. In such cases, the S-matrix needs to be
{\it resolved}. There is a fairly general procedure due to Fuchs,
Schellekens and Schweigert which one uses to obtain a resolved S-matrix
which has its usual properties\cite{resolve}.  Sometimes, however there
exists some discrete symmetry which distinguishes representations which
have the same character. In these cases, one can use the charge under the
discrete symmetry to obtain a resolved (or at least a partially resolved)
S-matrix. We refer the reader to ref. \cite{resolve} for the procedure to
resolve the S-matrix. In the case of Gepner models, we will discover that
this is the case generically and we will need to resolve the S-matrix
before using Cardy's solution to eqn. (\ref{consist}). 

\subsection{Brief review of Gepner models} 

Gepner models are exactly solvable supersymmetric compactifications of
type II string theory, where the internal part of the SCFT is constructed
by tensoring together $N=2$ minimal models. The central charge of the
minimal model of level $k$ is given by
\begin{equation}
c=\frac{3k}{k+2}
\end{equation}
A simple construction of the minimal model of level $k$ is realised by
adding one free boson to the $Z_k$ parafermionic field theory. This is
done as follows: from the free bosonic theory with the field denoted by
$\phi$, and the $Z_k$ parafermionic theory with parafermionic fields
labelled by $\psi_1$ and its hermitian conjugate, $\psi_1^{\dagger}$, one
can construct
\begin{eqnarray}
G(z)&=&\sqrt{\frac{2k}{k+2}}\psi_1:e^{i \phi \sqrt{\frac{k+2}{k}}}:
\nonumber \\
\overline{G}(z)&=&\sqrt{\frac{2k}{k+2}}\psi_1^{\dagger}
:e^{- i \phi \sqrt{\frac{k+2}{k}}}: \nonumber \\
J&=&i \sqrt{\frac{2k}{k+2}} \partial_z \phi
\end{eqnarray}
The operator product expansions for these generators satisfy the $N=2$
super conformal algebra.  The primary fields of the theory are labelled by
three integers $l,m,s$, and denoted by $\Phi^l_{m,s}$ whose dimension $h$
and $U(1)$ charge $q$ are given by are given by
\begin{eqnarray}
h&=&\frac{l(l+2) - m^2}{4(k+2)}~+~\frac{s^2}{8}
\nonumber \\
q&=&\frac{m}{k+2} - \frac{s}{2}
\end{eqnarray}
where $l=0,1, \cdots,k$ and $m=-(k+1),-k,\cdots,(k+2)$ mod $(2k+4)$
and $s=0,2,\pm1$.  The labelling integers satisfy the constraint $l+m+s\in 2Z$. 
In addition, there is an identification given by $(l,m,s)\sim
(k-l,m+k+2,s+2)$.  The $N=2$ characters of the
minimal models are defined in terms of the usual Jacobi theta functions
as: 
\begin{equation}
\chi_m^{l(s)}\left(\tau, z, u \right) = \sum_{j~{\rm mod}~k} C^l_{m+4j-s}
(\tau) \theta_{2m+(4j-s)(k+2),2k(k+2)}\left(\tau, 2kz, u \right)
\end{equation}
where $\theta_{n,m}(\tau,z,u)$ denotes the Jacobi theta function, and the
$C^l_m(\tau)$ are the characters of the parafermionic field theory. The
characters $\chi_m^{l(s)}$ have the property that they are invariant under $s
\rightarrow s+4$ and $m \rightarrow m+2(k+2)$ and are zero if $l+m+s \neq
0$ mod $2$. By using the properties of the theta functions, the modular
transformation of the minimal model characters is found to be
\begin{equation}
\chi_{m}^{l(s)} \left( -{1 \over \tau},0,0 \right) = C \sum_{l',m',s'}
\sin
\left({\pi(l+1)(l'+1) \over k+2} \right)\exp \left({i \pi mm' \over
k+2}\right)
\exp\left(- {i \pi ss' \over 2} \right) \chi_{m'}^{l'(s')}(\tau,0,0)
\end{equation}
where in the above sum one imposes $l'+m'+s'=0$ mod $2$ and $C$ is a
constant.

Gepner constructed compactifications of the heterotic string which had
spacetime supersymmetry by representing the internal part by a tensor
product of $N=2$ minimal models. His considerations are equally applicable
for compactifications of the type II string. Consider the tensor product
of $n$ minimal models of level $k_i$ ($i=1,\cdots,n)$. The total central
charge of the internal model is given by
\begin{equation}
c_{\rm int} =  \sum_{i=1}^{n} \frac{3k_i}{k_i+2}\quad,
\end{equation}
where $c_{\rm int}=15 -3d/2$, where $d$ is the dimensionality of
spacetime. Thus, for $d=4$, $c_{\rm int}=9$. Gepner constructs a spacetime
supersymmetric partition function by first projecting onto states for
which total $U(1)$ charges in both the left-moving and right-moving
sectors is an odd integer. Then, in order to preserve $N=1$ worldsheet
supersymmetry, the NS sector states of each sub-theory are coupled to each
other and do not mix with the $R$ sector states. He thus multiplies all
the $NS$ sector partition functions in each sub-theory and similarly for
other sectors i.e, $\widetilde{NS}$, $R$ and $\widetilde{R}$. 
Here, $NS$ refers
to the Virasoro character in the NS sector ($NS={\rm tr}_{NS}\
q^{L_0-c/24}$) while $R$ refers to the Virasoro character in the R sector
($R={\rm tr}_{R}\ q^{L_0-c/24}$). $\widetilde{NS}$ and $\widetilde{R}$
refer to the Virasoro characters in the appropriate sector with the
inclusion of $(-)^F$, where $F$ is the worldsheet fermion number
($\widetilde{NS}={\rm tr}_{NS}\ (-)^F\ q^{L_0-c/24}$). 
The full
partition function is a sum of the contributions from the four sectors.
Modular invariance of the full partition function is a consequence of
modular invariance in each of the sub-theories.

A related construction due to Eguchi et al.  makes use of {\it
supersymmetric characters} in order to construct modular invariant
partition functions for Gepner models\cite{eguri}.
The analysis of Gepner showed the relationship between
spacetime supersymmetry and spectral flow with $\eta={1\over2}$ in the
$N=2$ supersymmetry algebra.  The supersymmetric character is obtained by
first constructing the Virasoro character in the Neveu-Schwarz (NS) sector
and then including all characters (whose states are related to the
original one by spectral flow in steps of $\eta={1\over2}$). For example,
the graviton character is obtained by first considering the identity
operator.  Then, one applies the spectral flow operation once to obtain a
state in the Ramond sector.  The second application leads one back to the
NS sector. This procedure is repeated until one returns to the original
state after a few iterations. The supersymmetric character (in the
lightcone gauge) can be written as
\begin{equation}
X_i = {1\over2}\left\{NS_i \left({\theta_3\over\eta}\right)^m
-\widetilde{NS}_i \left({\theta_4\over\eta}\right)^m 
 - R_i \left({\theta_2\over\eta}\right)^m
+\widetilde{R}_i\left({\theta_1\over\eta}\right)^m\right\}\quad,
\end{equation}
where $({\theta\over\eta})^m$ come from level one SO$(2m)$ characters
associated with the non-compact spacetime of dimension $d$(with
$m=(d-2)/2$). The signs reflect the GSO projection required in order to
obtain the correct spin-statistics connection. 
As a consequence of
spacetime supersymmetry, each  supersymmetric character vanishes
identically. See ref. \cite{ictpgepner,gepner} for the details of the
argument. 

The modular invariant 
partition function on a torus for a type II string compactified on a
Gepner model is now constructed as follows.
One first constructs the supersymmetric character $X_0$ associated with
the graviton (this is associated with the identity operator in the Gepner
model).  One then obtains all other characters $X_i$ ($i=1,\cdots,r)$
which are obtained by applying the $S:\tau\rightarrow -{{1}\over\tau}$
modular transformation to $X_0$, the graviton character.  Then, one
constructs a modular invariant bilinear combination from the full set of
characters thus obtained. In the sequel, we will restrict ourselves to the 
case where the
partition function on the torus (for the type II string compactified on a
Calabi-Yau given by a Gepner model) is given by the following modular
invariant combination
\begin{equation} 
{\cal T} = \sum_{i=0}^{r}\ D_i \ |X_i|^2\quad, 
\end{equation} 
where $D_i={S_{0i}\over S_{i0}}$ is the multiplicity with which character
$X_i$ occurs in the torus partition function.

\subsection{Landau-Ginzburg description of Gepner models}

There is a lot of evidence that the level $k$ $N=2$ minimal model can be
obtained as the RG fixed point of a Landau-Ginzburg model (with global
$N=2$ supersymmetry)of a single scalar superfield with superpotential
$\Phi^{k+2}$. It has been shown that the central charge of the RG fixed
point matches that of the minimal model and more recently, the elliptic
genus of the two theories was shown to match\cite{wittenindex}. 

The massless spectra and symmetries of certain Gepner models are in one to
one correspondence with those obtained in some Calabi-Yau
compactifications\cite{ictpgepner,gepner}. This result was initially shown
by Gepner for the quintic hypersurface in $CP^4$ which is equivalent to
the $(k=3)^5$ Gepner model. For this example, it was shown in ref.
\cite{distler} that certain Yukawa couplings between the massless fields
also agreed from both sides. The explanation of this phenomenon came first
by a path integral argument due to Greene et. al \cite{gvw}.  Using the
relationship between the level $k$ $N=2$ minimal model and the the LG
theory with superpotential $\Phi^{k+2}$, for the Gepner model given by
$(k_1 ,k_2 ,...k_n)$, they chose the LG superpotential $W(\Phi_1, \Phi_2,
\cdots\Phi_n)=\Phi_1^{k_1+2}+\Phi_2^{k_2+2}+\cdots\Phi_n^{k_n+2}$.
Assuming that the $D$ terms in the theory are irrelevant operators and
their effect can be neglected in the path integral for this model, it was
shown in ref. \cite{gvw} that one exactly ends up with the constraint that
defines a Calabi-Yau manifold in weighted projective space.  There was a
need to impose a discrete identification in order to make the argument
work. This corresponds to an orbifolding of the LG model and gives rise to
the integer projection imposed by Gepner in order to have spacetime
supersymmetry. Thus the precise statement is that the Gepner model is
equivalent to the LG orbifold. The Calabi-Yau - Landau-Ginzburg
correspondence was later proved more rigourously by Witten
\cite{wittenphases} where it was shown how a varying K\"ahler parameter
interpolates between the geometrical (Calabi-Yau) and the non-geometrical
(Landau-Ginzburg) phases. 

For instance the string vacuum that corresponds to five copies of the
$k=3$ $N=2$ minimal model, is obtained by orbifolding by $\exp[i2\pi
J_0]$, where $J_0$ measures the left $U(1)$ charge. Other more complicated
orbifolding possibilities exist (and lead to other Calabi-Yau manifolds)
but we will not need to consider other possibilities in this paper.  A
$N=2$ LG theory which has not been orbifoldized contains only $(c,c)$ and
$(a,a)$ states. However, in order that a LG description of a $N=2$ super
conformal field theory reproduce the string vacuum it is essential that it
also include the $(a,c)$ states.  These states appear in the twisted
sector of the LG orbifold\cite{vafa,intril}.

\section{D-branes in Landau-Ginzburg models }

In this section, we will describe D - branes wrapped on supersymmetric
cycles using the Landau-Ginzburg description of Calabi-Yau manifolds. We
will first generalise the bulk Landau-Ginzburg theory by including
boundary terms which preserve part of the worldsheet supersymmetry
following the work of Warner\cite{warner}.  We will obtain the analog of A
and B type boundary conditions in this system. For the case of the
quintic, we will show that A-type boundary conditions naturally choose a
real submanifold which is the supersymmetric three-cycle constructed by
Becker et al.\cite{bbs}. 

We will consider the massive Euclidean Landau-Ginzburg theory in two
dimensions, with complex bosons $\phi_i$ and complex Dirac fermions
denoted by $\psi, \overline{\psi}$, with the left and right moving
components denoted by the subscripts $+$ and $-$ respectively.  The action
for the model (in which we have taken the boundary to lie on the line
$x^0\equiv x =0$ and $x^1\equiv y$) is given by
\begin{equation}
S= S_{\rm bulk} + S_{\rm boundary}\quad,
\end{equation}
where
\begin{eqnarray}
S_{\rm bulk} = \int_{- \infty}^{0}dx^0 \int _{- \infty}^{\infty} dx^1 
&&\left\{ -\left( \partial_\alpha\overline{\phi}_i \partial_\alpha
\phi_i \right)\right.\nonumber \\
&-&{1\over2} \left( \overline{\psi}_{-i}
\partial_0 \psi_{-i} - \overline{\psi}_{+i} \partial_0 \psi_{+i}
 - \partial_0(\overline{\psi}_{-i})\psi_{-i} + (\partial_0
\overline{\psi}_{+i})\psi_{+i}\right) \nonumber \\
&+& {i \over 2} \left(
\overline{\psi}_{-i}\partial_1 \psi_{-i} + \overline{\psi}_{+i} \partial_1 
\psi_{+i} - \partial_1(\overline{\psi}_{-i})\psi_{-i} - \partial_1(\overline
{\psi}_{+i})\psi_{+i}\right) \nonumber \\ 
&+&\left. \left( \frac{\partial^2 W}{\partial \phi_i \partial \phi_j}\right) 
\psi_{+i} \psi_{-j}  + \left( 
\frac{\partial^2\overline{W}}{\partial\overline{\phi}_i
\partial\overline{\phi}_j}\right)
\overline{\psi}_{+i} \overline{\psi}_{-j} -  |\frac{\partial W}{\partial 
\phi_i} |^2 \right\}  \nonumber \\
S_{\rm boundary}= 
\int_{- \infty}^{\infty}dy &&\left( -{1\over2}\overline{\psi}{_i} 
\gamma^* \psi_i \right)  
\end{eqnarray}

In the above $W(\phi)$ is a quasi-homogeneous superpotential. As is usual
for theories with boundary, the kinetic energy term for the fermions
written in symmetric form. In addition, we have included an explicit
boundary term following the work of Warner\cite{warner}.\footnote{The 
dictionary which relates Warner's notation to ours is as
follows: $ \lambda_1  =  \psi_{+i},\ \lambda_2  =  \psi_{-i},\ 
\overline{\lambda}_1  =  \overline{\psi}_{-i},\
\overline{\lambda}_2  =  \overline{\psi}_{+i},\
\alpha_1  =  \overline{\epsilon}_-,\
\alpha_2  =  \overline{\epsilon}_+,\
\overline{\alpha}_1  =  \epsilon_+,\
\overline{\alpha}_2  =  \epsilon_- $.}.
We have used an off diagonal basis where the two dimensional $\gamma$
matrices are defined by
\begin{equation}
\gamma^0 = \pmatrix{0&i\cr -i&0}\quad
\gamma^1 = \pmatrix{0&1\cr 1&0}\quad
\gamma^* = \pmatrix{1&0\cr 0&-1}\quad.
\end{equation}

The supersymmetry transformations of this model are given explicitly
in terms of the components to be
\begin{eqnarray}
\delta \phi_i &=& - \left( \psi_{+i} \epsilon_+ +  \psi_{-i}
\epsilon_- \right) \nonumber \\
\delta \overline{\phi}_i &=& \left( \overline{\psi}_{-i} \overline{\epsilon}
_- +  \overline{\psi}_{+i} \overline{\epsilon}_+ \right) \nonumber \\
\delta \psi_{+i}&=& \left( -\partial_0 \phi_i + i \partial_1 
\phi_i \right) \overline{\epsilon}_+ + \frac{\partial 
\overline{W}}{\partial \overline{\phi}_i}
\epsilon_- \nonumber \\
\delta \overline{\psi}_{+i}&=& \left( \partial_0 \overline{\phi}_i - i 
\partial_1 \overline{\phi}_i \right) \epsilon_+ 
 - \frac{\partial W}{\partial \phi_i} \overline{\epsilon}_-\nonumber \\
\delta \psi_{-i}&=& \left( \partial_0 \phi_i + 
i \partial_1 \phi_i \right) \overline{\epsilon}_- - \frac{
\partial \overline{W}}{\partial \overline{\phi}_i}\epsilon_+ \nonumber \\
\delta \overline{\psi}_{-i}&=& \left( - \partial_0 \overline{\phi}_i - 
i \partial_1 \overline{\phi}_i \right) \epsilon_- + \frac{\partial W}
{\partial \phi_i} \overline{\epsilon}_+
\end{eqnarray}
This action is now varied under ordinary and supersymmetric variation,
giving rise to boundary terms, and consistent boundary conditions are
imposed in order to cancel these. The boundary terms coming from ordinary
variation can be written as
\begin{eqnarray}
\delta_{ord}S =  - \int _{-\infty}^{\infty}dy (\partial_0
\overline{\phi}_i)\delta \phi_i&+&(\partial_0 \phi_i) \delta
\overline{\phi}_i
 + {1 \over 2} \left( \overline{\psi}_{-i} - \overline{\psi}_{+i}\right)
\left(\delta \psi_{+i} + \delta \psi_{-i} \right) \nonumber \\
&-& {1\over2}\left( \psi_{+i} - \psi_{-i}\right)\left(\delta
\overline{\psi}_{-i} + \delta \overline{\psi}_{+i} \right)
\end{eqnarray}
evaluated on the line $x=0$. Similarly, the boundary terms arising
out of supersymmetric variations of the action can be written as
\begin{eqnarray}
\delta_{susy}S = \int_{-\infty}^{\infty}dy
&&\left[-{1 \over 2}\partial_0
\phi_i\left(\overline{\psi}_{-i} - \overline{
\psi}_{+i}\right) + {i \over 2} \partial_1 \phi_i
\left( \overline{\psi}_{-i} + \overline{ \psi}_{+i}\right)\right]
\left(\overline{\epsilon}_- - \overline{\epsilon}_+ \right) \nonumber \\
&+&\left[ {1 \over 2}\partial_0 \overline{\phi}_i \left( \psi_{+i} - 
\psi_{-i}\right)  + {i \over 2}\partial_1\overline{\phi}_i
\left( \psi_{+i} +  \psi_{-i}\right)\right]\left(\epsilon_+ - 
\epsilon_- \right) \nonumber \\
&+&{1 \over 2} \left[ \left( \frac{\partial W}{\partial \phi_i}
\right) (\psi_{+i} + \psi_{-i})(\overline{\epsilon}_- + 
\overline{\epsilon}_+) - \left( \frac {\partial \overline{W}}
{\partial \overline{\phi}_i} \right) ( \overline{\psi}_{-i}
 + \overline{\psi}_{+i})(\epsilon_+ + \epsilon_-) \right] 
\end{eqnarray}

\subsection{A-type boundary conditions}

Following our earlier discussion on the A-type boundary conditions, we
will look for the unbroken $N=2$ supersymmetry to be given by
\footnote{One can also choose $\epsilon_+=-\overline{\epsilon}_-$ here.}
\begin{equation}
\epsilon_+=\overline{\epsilon}_- \quad.
\end{equation}
The above choice is dictated by A-type boundary condition $G_L^+ = \pm
G_R^-$ for the supersymmetry generators. 

The supersymmetric variation the action $S$ after imposing
$\epsilon_+=\overline{\epsilon}_-$ is
\begin{eqnarray}
\delta_{\rm susy} S = 
\int_{-\infty}^{\infty}dy
&&\left[-{1 \over 2}\partial_0
\phi_i\left(\overline{\psi}_{-i} - \overline{
\psi}_{+i}\right) + {i \over 2} \partial_1 \phi_i
\left( \overline{\psi}_{-i} + \overline{ \psi}_{+i}\right)\right]
 (\epsilon_+ - \epsilon_-)\nonumber \\
&+&\left[ {1 \over 2}\partial_0 \overline{\phi}_i \left( \psi_{+i} - 
\psi_{-i}\right)  + {i \over 2}\partial_1\overline{\phi}_i
\left( \psi_{+i} +  \psi_{-i}\right)\right]
(\epsilon_+ - \epsilon_-)\nonumber \\
&+&{1 \over 2} \left[ 
 \frac{\partial W}{\partial \phi_i}
 ( \psi_{+i}
 + {\psi}_{-i})
- \frac {\partial \overline{W}} 
{\partial \overline{\phi}_i} ( \overline{\psi}_{+i}
 + \overline{\psi}_{-i})
\right] (\epsilon_+ + \epsilon_-) 
\end{eqnarray}

Further, let us assume that the fermions also satisfy the following
condition:\footnote{Since $J_+=-J_-$ for A-type boundary conditions, we
are not permitted to set $\psi_{+i}+\psi_{-i}=0$ on the boundary. Thus one
has to choose $(\psi_{+i}-\overline{\psi}_{-i}) =0$ on the boundary.}
\begin{equation}
(\psi_{+i} - \overline{\psi}_{-i})|_{x=0} =0 
\label{bou1}
\end{equation}

The following set of boundary conditions on the bosonic fields makes the
action invariant under the $N=2$ supersymmetry.  The bosonic boundary
conditions are also consistent with the supersymmetric variation of the
fermionic boundary condition in eqn. (\ref{bou1}). 
\begin{eqnarray}
\partial_0 \left( \phi_i - \overline{\phi}_i \right)|_{x=0}&=& 0
\nonumber \\
\partial_1 \left(\phi_i + \overline{\phi}_i \right)|_{x=0}&=&0  \nonumber \\
\left. \left(\frac{\partial W}{\partial \phi_i} - \frac{
\partial \overline{W}}{\partial \overline{\phi}_i}\right)\right|_{x=0}&=&0
\label{bou2}
\end{eqnarray}
Hence (\ref{bou1}) and (\ref{bou2}) give us a set of boundary conditions
on the fields such that we have an unbroken $N=2$ supersymmetry on the
boundary. The last line of the eqn. (\ref{bou2}) has to be viewed as
a consistency condition on the boundary condition. It has a simple
interpretation (in the infrared limit) provided the equation $W=0$ admits
a pure imaginary solution. It corresponds to the statement that for 
directions along the brane, the variation of $W$ has to vanish. For example, 
for a
circle given by $f(x,y)=(x^2+y^2-1)=0$, the analogous statement is that
$\partial_\phi f =0$, where $\phi$ is the angle in cylindrical polar
coordinates. We will see that similar conditions appear even for B-type
boundary conditions whenever a Neumann boundary condition is imposed on
fields.

These `mixed' boundary conditions should correspond to a D- brane wrapped
on some cycle of Calabi-Yau given by the equation $W(\phi)=0$. Let us see
if this can be substantiated.  Notice that, the last of the equations in
(\ref{bou2}) implies that the real part of all the complex scalar
fields $\phi_i$ can be chosen to vanish on the boundary at $x=0$. Thus,
the target space interpretation is that the cycle corresponds to a 
submanifold of the Calabi-Yau given by the coordinates becoming
imaginary  on
the boundary.  As shown in \cite{bbs}, for the quintic hypersurface
defined in $CP^4$ by the equation
\begin{equation}
\Sigma_{i=1}^{5}\left( \phi_i \right)^5~=~0\quad,
\end{equation}
imposing the reality (or equivalently pure imaginary)\footnote{We will
nevertheless refer to this as real submanifold.}
condition on all the $\phi_i$ indeed provides one
with a submanifold which is a supersymmetric three-cycle.

Actually, (\ref{bou1}) and (\ref{bou2}) are not the most general 
choice of boundary conditions. The following set of boundary conditions is
more general: 
\begin{eqnarray}
(\psi_{+i} - {A_i}^j\ \overline{\psi}_{-j})|_{x=0} &=&0\nonumber  \\
\partial_0 \left( \phi_i -{A_i}^j\ \overline{\phi}_j \right)|_{x=0}&=& 0
\nonumber \\
\partial_1 \left(\phi_i +{A_i}^j\ \overline{\phi}_j \right)|_{x=0}&=&0 \nonumber \\ 
\left.\left({A_i}^j\ \frac{\partial W}{\partial \phi_j} - \frac{
\partial \overline{W}}{\partial \overline{\phi}_i}\right)\right|_{x=0} &=&
0\quad,
\end{eqnarray}
where the symmetric matrix $A$ satisfies $AA^{*}=1$ and it is block
diagonal i.e., it does not mix fields with different charge under the
$U(1)$ of the unbroken $N=2$ supersymmetry algebra. One simple choice is
given by $A={\rm Diag}(e^{i\theta_1},\cdots,e^{i\theta_n})$  subject to
the condition involving the superpotential being satisfied.

Given a matrix $A$ which provides boundary conditions consistent with the
superpotential, we can construct other consistent choices. Let us assume
that the superpotential is invariant under a discrete group $G$ which acts
holomorphically on the fields. Let $^g\phi_i={g_i}^j\phi^j$ be the the
action of $g\in G$. The invariance of the superpotential under $G$ implies
that $W(\phi)=W(^g\phi)$. Corresponding to the element $g$, we can
construct another $N=2$ preserving boundary condition on the fields given
by $A_g=g^{-1}\cdot A \cdot g^*$. Clearly, if $g$ is a real matrix, then
$A$ and $A_g$ belong to the same conjugacy class and we do not obtain new
boundary conditions.

Clearly with a LG theory it would be difficult to provide a description of
the boundary states in the cylinder channel with the same degree of
explicitness that we can associate with free-field theories. However we
can notice the following.  We can label the boundary states by the primary
fields associated with them as in the general case discussed by Cardy and
implemented by Recknagel and Schomerus.  Since for the A-type boundary
condition, one needs equal charges from the left and right moving sectors
in the construction of the boundary state, it is clear that the lowest
states are associated with the application of the LG fields themselves on
the ground state vacuum of the theory. It is clear that this may involve
appropriate number of $\phi$ fields, such that the $U(1)$ charge
projection condition is satisfied, a similar set of states with the
application of $\bar\phi$ fields and also states built by application of
both $\phi$ and $\bar\phi$ fields such that they have integral $U(1)$
charge. Some of these states will be obviously in the massive sector and
will not contribute to massless states but as we shall see later such
states are required in the general definition of the boundary state. This
ties in rather nicely with the method for the construction of boundary
states that we will pursue in section IV of the paper. In this connection
we note also that so far we have no means yet, strictly within the LG
formulation, to determine the normalization of the boundary states as is
done by the method of Cardy for the boundary states of an arbitrary
minimal model. 

\subsection{B-type boundary conditions}

Again, following the earlier analysis, for B-type boundary conditions
the $N=2$ supersymmetry is given by
\begin{equation}
\epsilon_+ = -\epsilon_- \quad.
\end{equation}
We will now look for boundary conditions on the fields such that the 
above supersymmetry is preserved. 

Under supersymmetry variation of the action, (after setting $\epsilon_+ =
-\epsilon_-$ as required), we obtain a boundary term of the form
\begin{eqnarray} \int_{-\infty}^{\infty}dy &&\left\{\left[\partial_0
\phi_i\left(\overline{\psi}_{-i} - \overline{ \psi}_{+i}\right) - i
\partial_1 \phi_i \left( \overline{\psi}_{-i} + \overline{
\psi}_{+i}\right)\right] \overline{\epsilon}_+ \right.\nonumber \\
&+&\left.\left[ \partial_0 \overline{\phi}_i \left( \psi_{+i}~-~
\psi_{-i}\right) + i\partial_1\overline{\phi}_i \left( \psi_{+i} +
\psi_{-i}\right)\right]\epsilon_+ \right \} \end{eqnarray} The vanishing
of the above boundary term suggests two possible boundary conditions:
\begin{enumerate} \item $\partial_0 \phi_i|_{x=0} =0$ and $(\psi_{-i} +
\psi_{+i})|_{x=0}=0$. This corresponds to Neumann boundary conditions on
the field $\phi_i$ and its complex conjugate $\overline{\phi}_i$.
Consistency with supersymmetry imposes the additional condition
${{\partial W}\over{\partial\phi_i}}|_{x=0}=0$. Note that this is a 
condition in spacetime where it says that the tangential
derivative along the boundary vanishes. 
\item $\partial_1 \phi_i|_{x=0} =0$ and $(\psi_{-i} -
\psi_{+i})|_{x=0}=0$.  This corresponds to Dirichlet boundary conditions
on the field $\phi_i$ and its complex conjugate $\overline{\phi}_i$. 
\end{enumerate}
Since the above set of boundary conditions treat both the real and
imaginary parts of the complex scalar fields $\phi_i$ in identical
fashion, the cycle which is chosen by the boundary conditions will
correspond to a holomorphic submanifold of the Calabi-Yau. Thus the cycle
is a supersymmetric cycle. 

Again, one can construct a general boundary condition. It is specified by
a hermitian matrix $B$ which satisfies $B^2=1$ and is block diagonal i.e.,
it does not mix fields with different charge under the $U(1)$ of the
unbroken $N=2$ supersymmetry algebra. The general boundary condition is
given by
\begin{eqnarray}
(\psi_{+i} + {B_i}^j\psi_{-j})|_{x=0}=0\quad, \nonumber  \\
\partial_0(\phi_i + {B_i}^j\phi_j)|_{x=0}=0\quad, \nonumber  \\
 \partial_1(\phi_i - {B_i}^j\phi_j)|_{x=0}=0\quad, \nonumber  \\
\left.\left({{\partial W}\over{\partial\phi_i}} + {B^*_i}^j {{\partial
W}\over{\partial\phi_j}}\right)\right|_{x=0}=0 
\end{eqnarray}
Since $B$ squares to one, its eigenvalues are $\pm1$.  An eigenvector of
$B$ with eigenvalue of $+1$ corresponds to a Neumann boundary condition
and $-1$ corresponds to a Dirichlet boundary condition.

What would B-type boundary states look like with Dirichlet or Neumann
boundary conditions on the LG fields? 
 With Neumann or Dirichlet boundary conditions it is
easy to see that the $U(1)$ current obeys boundary conditions that require
all boundary states to have equal and opposite charges in the left and
right moving sectors. This implies that all the boundary states for such
cycles must come from the twisted sector in the LG theory. It is not
immediately clear what difference the Neumann and Dirichlet boundary
conditions would make since in the twisted sector the zero-mode of the LG
fields are no longer present. However it is nevertheless clear that the
even supersymmetric cycles are charged under the Ramond-Ramond ground
states of the twisted sector. Before we turn to specific examples we would
like to add that all the massless states could probably be constructed by an
extension of the method of Kachru and Witten\cite{kachruwitten} where they
used the cohomology of the $\bar Q_+$ charge to define the massless states
in the left-moving sector of a (2,2) compactification of the heterotic
string. In the case of D-branes, in the open string sector, we have only one
$L_0$ operator and two supercharges. It is clear that an extension of the
methods of ref. \cite{kachruwitten} will be possible\cite{future}. 

\subsection{Examples}

We have seen that A-type boundary
conditions in the LG model are specified by a matrix $A$ and similarly
by a matrix $B$ for B-type boundary conditions. The choices of these
boundary conditions is not arbitrary. 
One has to in particular ensure that the consistency conditions
involving the superpotential are satisfied. 
In all the examples that we consider, for B-type
boundary conditions, we are unable to
impose Neumann boundary conditions on all fields simultaneously.
This is illustrated by considering
a simple example involving one scalar field (like the LG model associated
with the $N=2$ minimal model at level $k$). The only consistent boundary
condition one can impose in this case is the Dirichlet one. 

\subsubsection{The $1^3$ model}

This model is described by the superpotential involving three scalar
fields given by $W= (\phi_1^3+\phi_2^3+\phi_3^3)$. A-type boundary
conditions pick out the submanifold (one-cycle) given by
$$
(x_1^3+x_2^3+x_3^3)=0\quad,
$$
where $x_i = {\rm Im} \phi_i$. The discrete symmetry group of this
superpotential is given by $G = (S_3\times (Z_3)^3 )/Z_3$\footnote{$S_3$
is the permutation group with three elements (here it permutes the three
fields), the three $Z_3$'s are generated by the action $\phi^i
\rightarrow \omega
\phi_i$ (for $i=1,2,3$). ($\omega$ is a non-trivial cube root of unity.)
The quotient $Z_3$ is the diagonal $Z_3$.}. Other supersymmetric cycles
which can be constructed from this  cycle are $(ix_1,i\omega^a x_2,
i\omega^b x_3)$, where $a$ and $b$ are integers satisfying $a+b=0$ mod
$3$\footnote{This condition comes from requiring that the discrete
symmetry generator commute with the supersymmetry generator.}. These
correspond to choosing $A={\rm Diag}(1,\omega^a,\omega^b)$.  Thus, we end up 
with three $Z_3$ related cycles corresponding to $a=b=0$, $a=1,b=2$ and
$a=2,b=1$ respectively. One can verify that the three one-cycles are
non-intersecting.

There exists another choice for $A$ given by $A_1={\rm
Diag}[1,1,\exp(i2\pi/3)]$, which leads to the one-cycle given by
$$
(x_1^3+x_2^3-x_3^3)=0\quad,
$$
where $x_i = {\rm Im} \phi_i$ ($i=1,2$), $x_3={\rm Im} (\exp(-i\pi/3)
\phi_3)$.   By studying the action of $S_3$ on this cycle, we will see
that this cycle is not chosen in the Gepner model construction. 

Earlier, we had imposed the condition $a+b=0$ mod $3$ in the matrix $A$. 
Relaxing this condition, we will get two more sets of one-cycles
corresponding to $a+b=1,2$ mod $3$. Within each set, the one-cycles are
non-intersecting. However, if one considers one-cycles from different
sets, they can intersect. For example, the one-cycle chosen by  
$A={\rm Diag}(1,1,1)$ intersects the one-cycle chosen by
$A={\rm Diag}(1,\omega,\omega)$ at the point $(0,1,1)\simeq
(0,\omega, \omega)$ in homogeneous coordinates. The cylinder amplitude
between these two states will not vanish since the two boundary states
do not preserve the same supersymmetry generators. Further, one expects
to see a tachyon in the open string spectrum.

\bigskip
For B-type boundary conditions, we find the following consistent
choices:
\begin{enumerate}
\item Choose $B={\rm Diag}(-1,-1,-1)$ which corresponds to Dirichlet
boundary conditions on all scalars.  Let $\phi_i=c_i$ where
$c_i$ are constants. Presumably, they will have to
satisfy $(c_1^3+c_2^3+c_3^3)=0$ given the infrared limit of the bulk
theory but this does not follow from the consistency conditions.
Clearly $(c_1,c_2,c_3)$ corresponds to a point (in homogeneous
coordinates) on the torus and corresponds to a supersymmetric zero-cycle.
\item For $B=\pmatrix{0&-1&0\cr-1&0&0\cr0&0&-1}$, the consistency
conditions imply that $\phi_1+\phi_2=0$ and $\phi_3=0$ with
$(\phi_1-\phi_2)$ being free i.e., satisfying Neumann boundary conditions. 
\end{enumerate}
We are unable to find choices for $B$ such that one obtains two Neumann
and one Dirichlet boundary condition in addition to the all Neumann case
which can be clearly ruled out by analysing the consistency condition
involving the superpotential.

\subsubsection{The $2^2$ model}

This model is described by the superpotential $W=\phi_1^4 + \phi_2^4 +
\phi_3^2$, where we have included a `trivial' quadratic piece.  For A-type
boundary conditions given by $A=1$, there are no real solutions. However,
choosing $A={\rm Diag}(1,i,1)$, one obtains the one-cycle given by the
equation
$$
x_1^4 - x_2^4 + x_3^2 =0\quad,
$$
where $x_1={\rm Im} \phi_1$ , $x_2 = {\rm Im}(\phi_2/\sqrt{i})$ and $x_3={\rm
Im}\phi_3$. This equation has solutions for real $x_i$. The discrete
symmetry group of this model is given by $G=(S_2\times (Z_4)^2\times Z_2
)/Z_4$. Choosing an element of $G$ given by $g=(i^a,i^b, (-)^c)$ with
$a+b+2c=0$ mod $4$. By following the procedure mentioned earlier we obtain
$A_g = {\rm Diag}[(-)^a,i(-)^b,1]$ which provides cycles related to
$A={\rm Diag}(1,i,1)$ by a $Z_2$ subgroup. 

There is another choice given by $A_2={\rm Diag}(1,1,-1)$, one obtains the
one-cycle given by the equation
$$
x_1^4 + x_2^4 - x_3^2 =0\quad,
$$
where $x_1={\rm Im} \phi_1$ , $x_2={\rm Im} \phi_2$ and $x_3={\rm Re}
\phi_3$.  Again, this one-cycle is not chosen by the Gepner model model
construction. This cycle is invariant under the $S_2$ exchange while the
first choice is not invariant. 

\subsubsection{The Quintic}

We have already seen the example of a real three-cycle obtained from the
A-type boundary conditions with $A=1$. The Quintic has a discrete symmetry
group $G=(S_5\times (Z_5)^5/Z_5)$. A subgroup is given by the $Z_5$
generated by
$$g:(\phi_1,\phi_2,\phi_3,\phi_4,\phi_5)\rightarrow
(\phi_1,\alpha \phi_2,\alpha^2 \phi_3,\alpha^3 \phi_4,\alpha^4
\phi_5)\quad,
$$
where $\alpha$ is any non-trivial fifth root of unity. This boundary
condition corresponds to a three-cycle of the quintic which is related to
the real three-cycle by the $Z_5$ transformation. It follows trivially
that this cycle is a special Lagrangian submanifold of the deformed quintic and
hence a supersymmetric cycle. It is clear that this procedure leads to the
construction of supersymmetric cycles. Considering the full group $G$, one
can generate $G$-related supersymmetric cycles by the choice $g={\rm
Diag}(1,\alpha^a,\alpha^b,\alpha^c,\alpha^d)$, where $a,b,c,d$ are
integers which satisfy $a+b+c+d=0$ mod $5$.

	For B-type boundary conditions, we find the following three
consistent choices for the matrix $B$: (i) $B_1={\rm
Diag}(-1,-1,-1,-1,-1)$; (ii) $B_2=\pmatrix{-\sigma_1&0\cr 0& -1_{3\times3}}$	
and (iii) $B_3=\pmatrix{-\sigma_1 & 0 & 0 \cr 0 & -\sigma_1 & 0 \cr 0 & 0 &
-1}$,
where $\sigma_1=\pmatrix{0&1 \cr 1 &0}$ is a Pauli matrix. The first
choice is the all Dirichlet one. The second choice has one Neumann and
four Dirichlet conditions and the last one has two Neumann and three Dirichlet
conditions on some linear combinations of the fields.

\subsubsection{The Conifold}

The deformed conifold is described by a non-compact Calabi-Yau associated
with the superpotential\cite{gm,ghoshal}
$$
W = \phi_1^2 + \phi_2^2 + \phi_3^2 + \phi_4^2 -{\mu\over{\phi_5}}\quad,
$$
where $\mu=0$ is the conifold limit and $\mu=|\mu|e^{i\phi}$ is complex.
Imposing A-type boundary conditions with $A={\rm
Diag}(1,1,1,1,e^{2i\phi})$ chooses the three-cycle given by the equation
($x_i={\rm Im}\phi_i$, $i=1,2,3,4$ and $x_5={\rm Im}(\phi_5 e^{-i\phi})$)
$$
x_1^2 + x_2^2 + x_3^2 + x_4^2 -{|\mu|\over{x_5}}=0\quad.
$$
Working in inhomogeneous coordinates $y_i = x_i\sqrt{x_5}$, we obtain an
$S^3$ of radius $\sqrt{|\mu|}$ which is known to be a special Lagrangian
submanifold of the conifold and hence is a supersymmetric
cycle\cite{hitchina,hitchinb,stenzel}. 

\section{D-branes in Gepner models}

In this section we will be constructing the boundary states associated
with cycles of a Calabi-Yau space which can be obtained as a Gepner model. 
The Calabi-Yau is specified by tensoring together $N=2$ minimal models and
truncating to states with integer charge under the $U(1)$ of the $N=2$
supersymmetry. 

The partition function on a torus for a type II string compactified on a
Calabi-Yau manifold given by a Gepner model is given by\cite{eguri} 
\begin{equation} 
{\cal T} = \sum_{i=0}^{r}\ D_i \ |X_i|^2\quad, 
\end{equation} 
where $D_i={S_{0i}\over S_{i0}}$ is the multiplicity with which the
supersymmetric character $X_i$ occurs in the torus partition function 
and $X_0$ is the graviton character.

Since the multiplicities $D_i$ are generically not equal to one, one needs
to resolve the S-matrix associated with the Gepner model. There is a
procedure due to Fuchs, Schellekens and Schweigert which we employ to
resolve the S-matrix\cite{resolve}.  The Cardy prescription can then be
applied to the resolved S-matrix in order to obtain the boundary states
corresponding to D-branes wrapped around cycles of the Calabi-Yau
corresponding to the Gepner model. The resolution of the S-matrix for
models such as the quintic is computationally complex and hence we will
illustrate the procedure for the simple case of the $1^3$ and $2^2$ Gepner
models (for A-type boundary conditions). Here, we will see a very nice
match with respect to the analysis using the LG description and hence be
able to directly achieve a target space interpretation for the boundary
states. 

We should point out the differences between our approach and that of
Recknagel and Schomerus. In their construction, the boundary conditions
such as $J_+=J_-$ are imposed separately in each of the minimal models
which enters the theory after which they construct boundary states for by
tensoring together boundary states of the individual minimal models. Thus,
the boundary is forced to preserve the $N=2$ algebra of each minimal model
rather than the diagonal $N=2$. This seems to ensure that the setting is
``rational''. In our construction, we work with spectral flow invariant
orbits. Given the intimate relationship between spacetime supersymmetry
and spectral flow, our restriction may seem natural in the context of
D-branes since they are BPS states in spacetime. The supersymmetric
characters can be seen to be sums of characters of the extended algebra
${\cal W}$, one obtains by including the $\eta={1\over2}$ spectral flow
operator to the $N=2$ algebra\cite{eguri}. Thus, our boundary states
preserve the extended algebra ${\cal W}$ rather than the $N=2$ of the
individual minimal models. ``Rationality'' is obtained because we work
with only a finite number of supersymmetric characters rather than
characters of the irreducible representations of ${\cal W}$. We believe
that these two approaches complement each other and must not be considered
to be distinct.

\subsection{The $1^3$ Gepner model}

This $1^3$ Gepner model is obtained by the tensoring of three copies of
the $k=1$ $N=2$ minimal model. This is the Gepner model for a torus at its
SU$(3)$ point. The characters of the $k=1$ minimal model in the NS sector
will be labelled as follows.  ($\chi^l_m \equiv \chi^{l(s=0)}_{m} +
\chi^{l(s=2)}_{m}$)
\begin{center}
\begin{tabular}{|c|c|c|c|c|}\hline
$\chi^l_m$&q&h&Label\\[3pt] \hline
$\chi^{0}_{0}$&0&0&$A=\theta_{0,3}({\tau\over2})/\eta(\tau)$ \\[3pt] \hline
$\chi^{1}_{1}$&1/3&1/6&$B=\theta_{2,3}({\tau\over2})/\eta(\tau)$\\[3pt] \hline
$\chi^{1}_{-1}$&-1/3&1/6&$B_c=\theta_{4,3}({\tau\over2})/\eta(\tau)$\\[3pt] \hline
\end{tabular}
\end{center}
$B$ is associated with a chiral primary state and $B_c$ is associated with
an anti-chiral primary state. Under spectral flow(with $\eta=1$), we have
the sequence
$$
A\rightarrow B\rightarrow B_c\rightarrow A
$$ 

The spectral flow invariant orbits for this model in the NS sector are
\begin{center}
\begin{tabular}{|c|l|c|}\hline
Label & ~~~~~~~~Orbit & $q$; $h$ \\ \hline 
$NS_0$ & $A^3 +B^3 + B_c^3$ & $q=h=0$\\ \hline 
$NS_1$ & $3 A B B_c$ & $q=0$; $h={1\over3}$\\ \hline
\end{tabular}
\end{center}
In the above table, the values of $q$ and $h$ correspond to the state with
the smallest value of $h$ occurring in the spectral flow invariant $NS$
orbit.  $NS_0$ is the graviton orbit and the other orbit is massive i.e.,
it corresponds to massive states in the non-compact spacetime.  The choice
of $3 A B B_c$ rather than $ A B B_c$ as the character for the $NS_1$
state can be understood as follows: Let us assume that the three minimal
models are labelled $1,2,3$ respectively. Then a spectral flow invariant
orbit is given by $(A_1 B_2 B_{c3} + B_1 B_{c2} A_3 + B_{c1} A_2 B_3)$,
where we have explicitly kept the minimal model label. Getting rid of
these labels leads to $3 A B B_c$ and hence our choice. The S-matrix for
this model is derived to be
$$
S={1\over\sqrt{3}}\pmatrix{1&2\cr 1&-1}
$$
$D_0=1$ and $D_1=2$.  It is sufficient to consider the NS sector to obtain
the S-matrix. One can show that this S-matrix is identical to that
obtained from the modular transformation of the full supersymmetric
character\cite{ictpgepner,gepner}.  A modular invariant torus partition
function for this model is given by
\begin{equation}
{\cal T} = \sum_{i=0}^1\ D_i\ |X_i|^2
\end{equation}
where $X_i$ are the supersymmetric characters\footnote{The multiplicity of
two associated with $NS_1$ again is related to the fact that if we kept
track of the minimal model labels, there are two distinct spectral flow
invariant orbits given by the even permutation $NS_{1} =(A_1 B_2 B_{c3}
+ B_1 B_{c2} A_3 + B_{c1} A_2 B_3)$ and the odd permutation $NS_{1}^c=(B_1
A_2 B_{c3} + B_{c1} B_2 A_3+ A_1 B_{c2} B_3)$. This actually completely
resolves the S-matrix here. In more complicated situations, this will
enable us to partially resolve the S-matrix.}. 

However, as things stand one cannot apply Cardy's prescription here since
the character $X_1$ occurs with multiplicity $2$ in the toroidal partition
function.  In order to obtain the resolved S-matrix, one splits
$D_1=2=1+1$. Thus the resolved S-matrix is a $3\times3$ matrix. 
It is
\begin{equation} \widetilde{S} = {1\over\sqrt{3}}
\pmatrix{1&1&1\cr1&\omega&\omega^2\cr 1&\omega^2&\omega} \end{equation}
where $\omega$ is a cube root of unity.  
This
resolved S-matrix is the S-matrix for the SU$(3)$ level $k=1$
Wess-Zumino-Witten model which is consistent with the fact that this
Gepner model corresponds to compactification on a torus at its SU$(3)$ 
point. 
One can check that:
\begin{itemize} \item $\widetilde{S}^2 = C$ where
$C=\pmatrix{1&0&0\cr0&0&1\cr0&1&0}$ is the charge conjugation matrix.  
\item
$\widetilde{S}$ is symmetric and unitary. 
\item $\widetilde{S}^4=1$. 
\item $(\widetilde{S}T)^3 = \widetilde{S}^2$ with $T=$
Diag$(-i,-i\omega, -i\omega)$.
\end{itemize}

Let
$|0\rangle\rangle_A$, $|1\rangle\rangle_A$, $|1^c\rangle\rangle_A$ be
the Ishibashi states (associated with the characters $X_0$, $X_{1}$,
$X_{1}^c$) which satisfy A-type boundary conditions. 
We apply Cardy's procedure using the resolved S-matrix and obtain
the following boundary states
\begin{eqnarray}
|\tilde{0}\rangle &=& 3^{-1/4} \left (|0\rangle\rangle_A +
|1\rangle\rangle_A + |1^c\rangle\rangle_A \right)\\
|\tilde{1}\rangle &=& 3^{-1/4} \left (|0\rangle\rangle_A +
\omega |1\rangle\rangle_A +\omega^2 |1^c\rangle\rangle_A \right)\\
|\tilde{2}\rangle &=& 3^{-1/4} \left (|0\rangle\rangle_A +
\omega^2 |1\rangle\rangle_A +\omega |1^c\rangle\rangle_A \right)
\end{eqnarray}
Note that if we kept track of the minimal model labels, under the exchange
of labels $2\leftrightarrow3$, $ |1\rangle\rangle_A
\leftrightarrow|1^c\rangle\rangle_A$. Thus, under the same exchange
$|\tilde{1}\rangle \leftrightarrow |\tilde{2}\rangle$ with the state
$|\tilde{0}\rangle$ being invariant. These boundary states fit in
beautifully with the analysis of the $1^3$ model in the LG description.
There we obtained a set of A-type boundary conditions parametrised by
$A={\rm Diag}(1,\omega^a,\omega^b)$ with $a+b=0$ mod $3$. We make the
following correspondence: The $a=b=0$ boundary condition is identified
with the state $|\tilde{0}\rangle$ and $(a,b)=(1,2),(2,1)$ with the other
two boundary states (using properties of these b.c.'s under the
$2\leftrightarrow3$ exchange in the LG model)\footnote{The other choice of
boundary condition given by $A_1={\rm Diag}(1,1,\exp[i2\pi/3])$ is ruled
out because the equation for the one-cycle is clearly not invariant under
the $2\leftrightarrow3$ exchange.}

A more direct correspondence can be worked out by considering the part of
the boundary state involving only the $(c,c)$ and $(a,a)$ states.
Following the analysis in the LG orbifold, the boundary condition given by
the matrix $A={\rm Diag}(1,\omega^a,\omega^b)$ corresponds to multiplying
the $(a,a)$ field by the phases given in $A$. Let
$(\bar{\phi}_1,\bar{\phi}_2,\bar{\phi}_3)$ be the generators of the
$(a,a)$ ring. Multiplying this by $A$, one sees that $|0\rangle\rangle_A
\rightarrow |0\rangle\rangle_A $, $|1\rangle\rangle_A \rightarrow
\omega^a |1\rangle\rangle_A $ and $|1^c\rangle\rangle_A \rightarrow
\omega^b |1^c\rangle\rangle_A $, where we have used $a+b=0$ mod $3$. Thus,
these boundary states are related to D1-branes of the type IIB string
wrapping around non-intersecting supersymmetric one-cycles on the torus at
the SU$(3)$ point as follows from the LG analysis of the earlier
section.  We can also compare with the result of Recknagel and Schomerus. 
The nine states they obtain for this model can be grouped into sets
of three. The grouping is chosen by the condition that within each set,
the same spacetime supersymmetry is preserved by all states. Thus,
within a set, off-diagonal cylinder amplitudes should vanish by
supersymmetry. The three states we construct belong to one set.

It is easy to verify that the cylinder partition function ${\cal
C}_{\tilde{i}\tilde{i}} = 3^{-1/2} (X_0 +  X_1 + X_{1}^c)$. This reflects 
the $Z_3$
relationship between the three supersymmetric cycles. Under a modular
transformation, the annulus partition function we obtain is given by
${\cal A}_{\tilde{i}\tilde{i}} = X_0$. This implies that
$n_{\tilde{0}\tilde{0}}^i = \delta_{i0}$ i.e, only the identity sector
propagates in the vacuum channel. Both amplitudes vanish as required by
supersymmetry.  Finally, the annulus amplitude ${\cal
A}_{\tilde{0}\tilde{i}} = X_i$. Thus, we see that the massive character
$X_1$ is related to off-diagonal D-brane configurations (i.e., a D-brane
configuration that begins at one boundary and ends at another).

\subsection{The $2^2$ Gepner model}

This Gepner model describes a torus at the SU$(2)\times$SU$(2)$ point. The
characters of the $k=2$ minimal model in the NS sector will be labelled as
follows.  ($\chi^l_m \equiv \chi^{l(s=0)}_{m} + \chi^{l(s=2)}_{m}$)
\begin{center}
\begin{tabular}{|c|c|c|l|}\hline
$\chi^l_m$&q&h&~~~~~~~~~~Label\\[3pt] \hline
$\chi^{0}_{0}$&0&0&$A=\chi_0(\tau)\theta_{0,2}(2\tau)+
\chi_{1\over2}(\tau)\theta_{2,2}(2\tau)$\\[3pt] \hline
$\chi^{2}_{2}$&1/2&1/4&$B=\chi_0(\tau)\theta_{1,2}(2\tau)+
\chi_{1\over2}(\tau)\theta_{3 ,2}(2\tau)$\\[3pt] \hline
$\chi^{2}_{-2}$&-1/2&1/4&$B_c=\chi_0(\tau)\theta_{3,2}(2\tau)+
\chi_{1\over2}(\tau)\theta_{1,2}(2\tau)$\\[3pt] \hline
$\chi^{2}_{0}$&0&1/2&$C=\chi_0(\tau)\theta_{2,2}(2\tau)+
\chi_{1\over2}(\tau)\theta_{0,2}(2\tau)$\\[3pt] \hline
$\chi^{1}_{1}$&1/4&1/8&$D=\chi_{1\over{16}}(\tau)\theta_{1 ,2}(\tau)$
\\[3pt] \hline
$\chi^{1}_{-1}$&-1/4&1/8&$D_c=\chi_{1\over{16}}(\tau)
\theta_{3,2}(\tau)$\\[3pt] \hline
\end{tabular}
\end{center}
where 
\begin{eqnarray*}
\chi_0(\tau) &=& {1\over2}\left(\sqrt{{\theta_3(\tau)}\over{\eta(\tau)}} +
\sqrt{{\theta_4(\tau)}\over{\eta(\tau)}}\right)\quad,\\
\chi_{1\over2}(\tau) &=& {1\over2}\left(\sqrt{{\theta_3(\tau)}\over{\eta(\tau)}}
 - \sqrt{{\theta_4(\tau)}\over{\eta(\tau)}}\right)\quad{\rm and}\\
\chi_{1\over{16}}(\tau)&=&\sqrt{{\theta_2(\tau)}\over{2\eta(\tau)}}
\end{eqnarray*}
are the Ising model characters.
Under spectral flow (with $\eta=1$), we have the sequences 
$$
A\rightarrow B\rightarrow C\rightarrow B_c\rightarrow A
$$ 
$$
D\rightarrow D_c \rightarrow D
$$.

The spectral flow invariant orbits for this model are
\begin{center}
\begin{tabular}{|c|l|c|}\hline
Label & ~~~~~~~~Orbit & $q$; $h$\\ \hline
$NS_0$ & $A^2 +B^2 + C^2 + B_c^2$ & $q=h=0$\\  \hline
$NS_1$ & $2(A C + B B_c)$ &$q=0;h={1\over2}$\\ \hline
$NS_2$ & $ 2D D_c$& $q=0;h={1\over4}$\\ \hline
\end{tabular}
\end{center}

$NS_0$ is the graviton orbit and the other two orbits are massive. The
S-matrix for this model is derived to be 
\begin{equation}
S={1\over2}\pmatrix{1&1&2\cr 1&1&-2\cr 1&-1&0}
\end{equation}
$D_0=1$, $D_1=1$ and $D_2=2$.

In order to resolve the fixed point ambiguity, we need to split the $D_2$
as the sum of squares. $D_2$ can be written as $1+1$ leading to a
resolution of $S$ as a $4\times4$ matrix. The resolved S-matrix is given
by
\begin{equation}
\widetilde{S}={1\over2}\pmatrix{1&1&1&1\cr1&1&-1&-1\cr1&-1&-1&1\cr1&-1&1&-1}
\end{equation}
This resolved S-matrix is the S-matrix for the SO$(4)$ level $k=1$
Wess-Zumino-Witten model  which is again consistent with the symmetry
associated with the spacetime torus corresponding to this Gepner
model.
As one can see, $\widetilde{S}$ is symmetric and squares to the identity
matrix. 

Now, one can apply Cardy's procedure using the resolved S-matrix. Let
$|0\rangle\rangle_A$, $|1\rangle\rangle_A$
$|2^+\rangle\rangle_A$, $|2^-\rangle\rangle_A$ be the
Ishibashi states associated with the characters $X_0,X_1,X_{2,\pm}$
which satisfy A-type boundary conditions. Then Cardy's formula leads to
the following four boundary states:
\begin{eqnarray}
|\tilde{0}\rangle &=& 2^{-1/2} \left (|0\rangle\rangle_A +
|1\rangle\rangle_A+
|2^+\rangle\rangle_A + |2^-\rangle\rangle_A \right)\\
|\tilde{1}\rangle &=& 2^{-1/2} \left (|0\rangle\rangle_A
+|1\rangle\rangle_A -
|2^+\rangle\rangle_A - |2^-\rangle\rangle_A \right)\\
|\tilde{2}\rangle &=& 2^{-1/2} \left (|0\rangle\rangle_A
-|1\rangle\rangle_A -
|2^+\rangle\rangle_A + |2^-\rangle\rangle_A \right) \\
|\tilde{3}\rangle &=& 2^{-1/2} \left (|0\rangle\rangle_A
-|1\rangle\rangle_A +
|2^+\rangle\rangle_A - |2^-\rangle\rangle_A \right)
\end{eqnarray}
We thus obtain four boundary states. These four states are related to each
other by an $S_2\times Z_2$ subgroup of the discrete symmetry group. The
$Z_2$ is the same one which gave different one-cycles in the LG
description.  The boundary state $|\tilde{0}\rangle$ can be identified
with the boundary condition corresponding to $A={\rm Diag}(1,i,1)$. We
relegate to the appendix the detailed discussion as to how the other
choice for $A$ is ruled out. 

We will now compare with the results of Gutperle and Satoh (GS) for the
$2^2$ model obtained by using the method of Recknagel and Schomerus. One
can show that $NS_0 = \theta_3(\tau)[\theta_3^2(\tau) +
\theta_4^2(\tau)]/\eta(\tau)$ and $NS_1 = \theta_3(\tau)[\theta_3^2(\tau)
- \theta_4^2(\tau)]/\eta(\tau)$. (Here $\eta$ is the Dedekind eta function
and $\theta_i$ are the standard theta functions.) The annulus amplitude
${\cal A}_{\alpha\alpha}= X_0$ which can be seen to be equal to partition
function for $(l_1',l_2')=(0,0)$ in the notation of GS (upto factors of
$\eta$). The boundary state $|\tilde{0}\rangle + |\tilde{1}\rangle$ gives
rise to the annulus amplitude $(2X_0 + 2X_1)$ which is equal to the GS
calculation for $(l_1',l_2')=(1,1)$. Interestingly, there does not seem to
be a consistent boundary state which can give rise to the
$(l_1',l_2')=(1,0)$. For example, there is a state given by the
combination of Ishibashi states $|0\rangle\rangle_A + |1\rangle\rangle_A$
which cannot be written as an integer sum of the four boundary states we
have constructed. This state gives the annulus amplitude for
$(l_1',l_2')=(1,0)$ but is ruled out by its incompatibility with eqn.
(\ref{consist}). 

\subsection{The $1^6$ Gepner model}

In order to illustrate the increase in the degree of complexity, we
consider the simplest non-toroidal model: the $1^6$ Gepner model.  This
corresponds to one of the orbifold points in K3 moduli space. The notation
for the $k=1$ characters are as in the $1^3$ model. 
\begin{center}
\begin{tabular}{|c|l|c|}\hline
Label & ~~~~~~~~~~Orbit & Multiplicity \\ \hline
$NS_0$ & $A^6 +B^6 + B_c^6$ & 1 \\   \hline
$NS_1$ & $A^3 B^3 + B^3 B_c^3 + B_c^3 A^3$ &20  \\ \hline
$NS_2$ & $3A^2 B^2 B_c^2 $&30 \\  \hline
$NS_3$ & $A^4 B B_c + A B^4 B_c + A B  B_c^4$&30  \\  \hline
\end{tabular}
\end{center}
$NS_0$ corresponds to the graviton orbit, $NS_1$ is a massless orbit and
$NS_2$, $NS_3$ are massive orbits. In the above table, by multiplicity we
mean the number of distinct orbits which occur if we keep track of the
minimal model labels. 

The S-matrix is calculated from the S-matrix of the minimal model to be
$$
S={1\over{9}}\pmatrix{1&20&30&30\cr1&-7&3&3\cr1&2&3&-6\cr1&2&-6&3}
$$
$D_0=1$, $D_1=20$, $D_2=30$ and $D_3=30$. The resolved S-matrix is
expected to be an $81\times 81$ matrix which increases the complexity of
the operation. However, in this example, if one keeps track of the minimal
model labels, one should in principle be able to directly compute the
resolved S-matrix. This is because we find that the multiplicity is equal
to the $D_i$ associated with the orbit. This is not generically true. This
model is presumably tractable if one uses a computer program to automate
the process. 

\section{Discussion and Conclusions}

In this paper, we have studied D-branes wrapped around supersymmetric
cycles using boundary LG as well as boundary CFT formulations. The LG
formulation is suitable for understanding the boundary conditions from the
target space viewpoint while the boundary CFT formulation provides the
corresponding boundary state. The common discrete symmetry group
associated with both the LG orbifold and the corresponding Gepner model
has been a useful tool in relating boundary conditions to boundary states.
It also suggests that the boundary states constructed by Recknagel and
Schomerus by tensoring boundary states for the individual minimal models
may be further classified by means of charges associated with the discrete
symmetry group. In our method, this is also seen through the resolution of
the S-matrix of the Gepner model. 

  Clearly it is important to extend the program of studying closed string
vacua for Calabi-Yau compactifications involving the use of LG models and
the general structure of N=2 superconformal theories to the case of
D-brane states. In relation to the approach to this problem that we have
adopted in this paper the following points are worth noting: 

i) We need to extend the use of the Landau-Ginzburg model techniques so
that more relevant information can be extracted. As has been noted by
other authors, this may involve the extension of the methods of the $N=2$
topological field theory techniques to the case of boundary $N=2$ SCFTs.
In particular, index calculations of various kinds may be performed in the
LG model using purely free-field techniques by extension of similar
techniques used in the closed string case \cite{wittenindex}. For example,
it would be useful to compute the Tr$(-1)^F$ in the Ramond sector of the
open string by such techniques and compare them to the calculations of ref.
\cite{quintic}. 

ii) It is clear that some information on the boundary states can be
obtained even if they will not approach the level of detail that is
characteristic of the flat space case. In this connection, as we have
emphasised earlier, our construction of boundary states in the case of the
Gepner models uses a modular transformation matrix that acts on the
conformal blocks that are basically spectral flow invariant orbits. 
Clearly this construction carries a lesser amount of detail than the
boundary state construction of ref. \cite{reck} but could well make more
transparent the connection between the boundary states in the tensor
product of minimal models and the boundary states in the LG approach. 

iii) The construction that we have used here for the boundary states seems
a priori difficult to extend to the case of K3 and Calabi-Yau three-fold
compactifications. In particular the fixed point resolution would appear
to be hopelessly complicated even in the simplest cases. But since the
resolution would involve presumably no more than the use of the full
symmetry of the model it might be possible to solve the problem by
computer techniques. In such a situation, the results presented for the
$T^2$ in this paper would be extendable to the case of compactifications
like the quintic Calabi-Yau. The diagonal partition functions (that is
between identical branes) in the cylinder channel and hence in the annulus
channel are however known even despite the fixed point resolution even in
the complicated cases by our construction. Its extension to non-diagonal
cases by our methods would be of considerable interest. 

iv) It is of interest to see whether the LG-CY correspondence shown by
Witten by making use of linear sigma models will go through for the case
of linear sigma models with boundary\cite{jlos}. In this paper, we
introduced a generalised boundary condition parametrised by two matrices
$A$ and $B$ for A-type and B-type boundary conditions respectively. It
will be useful to examine these matrices in the context of the linear
sigma model.

\bigskip
\noindent{\bf Acknowledgments} We thank M. Blau, J. Majumdar, K. Paranjpe,
A. Sen, K. Srinivas and S. Wadia for useful discussions.  T. Sarkar would 
like to thank the Abdus Salam ICTP, Trieste for its hospitality, where part 
of this work was completed. T. Jayaraman would like to thank the Abdus Salam
ICTP, Trieste for hospitality at the Extended Workshop on String Theory
during the course of which this paper was finalised.

\appendix
\section{Discrete symmetries and the $2^2$ boundary states}

In this appendix, we will discuss how the discrete symmetries of the $2^2$
model enables us to understand the resolution of the S-matrix and make
connection with the choice of boundary condition in the $2^2$ LG model. 

The character $D$ is associated with the LG field $\phi$. We will thus use
$\phi_i$ to represent the corresponding chiral primary in the $i$-th
minimal model. Thus the character $B$ is associated with $\phi^2$. The
part of the Ishibashi state involving only the chiral primaries associated
with $DD_c$ will look something like
$$
|2^\pm\rangle\rangle = (\phi_1 \bar{\phi}_2 \pm \phi_2 \bar{\phi}_1)|0\rangle
+ \cdots\quad,
$$
where there is a sign ambiguity in the definition if we require that it be
an eigenstate of the permutation group $S_2$. Both states will be
associated with the same character $NS_2 = 2 D D_c$.  The resolution of
the S-matrix distinguishes between these two boundary states. Under $S_2$,
 we have that $|2^\pm \rangle \rangle\rightarrow \pm
|2^\pm\rangle\rangle$. The Ishibashi state associated with $NS_0$ remains
invariant under this $S_2$. However, for the character associated with
$NS_1$, there are two possible Ishibashi states
$$
|1^\pm\rangle\rangle = (\phi_1^2 \bar{\phi}_2^2 \pm \phi_2^2
\bar{\phi}_1^2)|0\rangle
+ \cdots\quad,
$$
where $\pm$ denotes the $S_2$ eigenvalue. Requiring that $S_2$ relate the
boundary state $|\tilde{0}\rangle$ to either $|\tilde{2}\rangle$ or
$|\tilde{3}\rangle$ picks the minus sign. Thus, we get under this $S_2$
$|\tilde{0}\rangle \leftrightarrow |\tilde{3}\rangle$ and
$|\tilde{1}\rangle \leftrightarrow |\tilde{2}\rangle$. 

There is another $Z_2$ subgroup of the discrete symmetry group generated
by $\phi_1\rightarrow i \phi_1$ and $\phi_2\rightarrow -i \phi_2$ (This
corresponds to $a=1$,$b=3$ using the notation given in the examples
section for the $2^2$ model.) One can check that under this $Z_2$, $|2^\pm
\rangle \rangle\rightarrow - |2^\pm\rangle\rangle$. One can also see that
the states associated with $NS_0$ and $NS_1$ remain invariant under this
$Z_2$. Under the action of this $Z_2$ one has $|\tilde{0}\rangle
\leftrightarrow |\tilde{1}\rangle$ and $|\tilde{2}\rangle \leftrightarrow
|\tilde{3}\rangle$

In order to translate this picture into the LG language let us summarise
the effect of the two discrete groups on the LG fields. Under the $S_2$,
$\phi_1\leftrightarrow\phi_2$ and under the $Z_2$, $\phi_1\rightarrow
i\phi_1$ and $\phi_2\rightarrow -i \phi_2$. We had discovered two
different boundary conditions in the LG model given by $A={\rm
Diag}(1,i,1)$ and $A_2={\rm Diag}(1,1,-1)$. Under the $S_2\times Z_2$, $A$
gives rise to four different boundary conditions, while the $A_2$ boundary
condition is invariant under $S_2$. Thus the Gepner model construction
seems to choose the  boundary condition specified by $A$.

\end{document}